# Discovery of Elliptic Curve Cryptographic Private Key in O(n)

*Abstract:* An algorithm is presented that in context of public key use of Elliptic Curve Cryptography allows discovery of the private key in worst case O(n).

Charles Sauerbier

December 2009

Humans have for as long as recorded history sought means to keep secrets and to communicate in secret. In modern times we have turned to electronic computations as means to facilitate both. The later problem of communicating in secret has had many interesting solutions proposed; some succeeded for a time; others failed – sometimes miserably. Counted midst their numbers is the system of Elliptic Curve Cryptography; a subject on which we will never profess to be an expert, but one that does fail miserably in practice.

**Elliptic Curve Cryptography**[1][2][3]

The ECC system is based on having defined several factors, described by the tuple (p, a, b, G, n, h). Added to these common factors are the individual's specific private key 'd' and public key 'Q'. Of the many factors one is kept as a secret – the private key 'd' – while all others are shared. The factors shared include the tuple of common factors that are specific to definition of the Abelian group on which the keys are premised. Of the shared factor one is the public key 'Q'.

The security of ECC rests on how difficult it is to find 'd' – private key – given all the shared factors, and specifically the public key 'Q'. The relationship between the private key and the public key is defined to be Q = d • G, where G – a shared common factor – is a well known base point. The operation • is defined as an arithmetic function that is fundamentally a difference equation; the consequence of which is to "walk" through a sequence of points. The specific points are described by an elliptic curve. The specifics of the curve are irrelevant to factoring the private key.

**Weakness in Scheme**

While the premise that factoring the Elliptic Curve is difficult may be valid, the problem is that to obtain a key pair (d, Q) one must perform some number of operations from some point of origin. One then, in the context of a fully observable algebraic structure provides half the key pair to others. However, the computation of the key pair is analogous to computing a Fibonacci number (by brute force without using the many tricks). Both the computation of the key pair and the pair (n, $F_n$) amount to the application of a difference equation[4] on an algebraic group. The Fibonacci numbers have the advantage of being a computationally uncountable set, while the field over which the ECC key pair is computed is a finite set of cardinality bounded by the value of 'p' in the tuple (p, a, b, G, n, h).

---

[1] (Washington, 2008)

[2] At time of writing subject related material could be found at: http://en.wikipedia.org/wiki/Elliptic_curve http://en.wikipedia.org/wiki/Elliptic_curve_cryptography, http://www.certicom.com/index.php/ecc-tutorial.

[3] The animation at time of writing at http://www.certicom.com/index.php/212-adding-the-points-p-and-p can be visually assistive in understanding the nature of the problem, as well as providing geometric validity of the weakness.

[4] (Goldberg, 1986) (Elaydi, 2005)

The security of ECC is premised on the fallacy that to determine the private key it is necessary to perform some inverse function on the public key and computation of that inverse is difficult. The reality is that discovering the private key is no more difficult that computing the public key.

**Factoring the Private Key**

Given some ECC scheme based on E = (p, a, b, G, n, h)[5] to find the private key it is only necessary to perform the iterative process $k = \bullet_{i=1}^{k=Q} G$, where • is the operator used to compute the public key, while counting the iterations necessary to reach k = Q.

**Conclusion**

While ECC could serve well as a symmetric key scheme its value as a public key scheme is questionable, as the difficulty of discovering the private key given the public key is O(n) – never being more difficult than computing the public key.

**References**

Elaydi, S. (2005). *An Introduction to Difference Equations, 3rd Ed.* Springer.

Goldberg, S. (1986). *Introduction to Difference Equations (Paperback Reprint).* Dover Publications.

---

[5] (m, f, a, b, G, n , h) in the binary instance of the scheme.